\documentclass[aps,prl,superscriptaddress,amsmath,amssymb,floatfix,twocolumn,showpacs]{revtex4}

\pdfoutput=1

\usepackage{amsmath,dcolumn,bm,amsthm,amsfonts,amssymb,tabularx,subfigure}
\usepackage[colorlinks=true,urlcolor=blue,citecolor=blue,linkcolor=blue]{hyperref}
\usepackage{times,bbold,dsfont}
\usepackage{graphicx,graphics,color}

\newcommand{\beq}{\begin{equation}}
\newcommand{\eeq}{\end{equation}}
\newcommand{\bea}{\begin{eqnarray}}
\newcommand{\eea}{\end{eqnarray}}
\newcommand{\bal}{\begin{align}}
\newcommand{\eal}{\end{align}}

\begin{document}

\title{Type-II Topological Meissner States}

\author{Chun~Chen}
\email[Corresponding author. \\]{cchen@physics.umn.edu}
\affiliation{School of Physics and Astronomy, University of Minnesota, Minneapolis, Minnesota 55455, USA}

\author{Wei~Yan}
\affiliation{Department of Physics and State Key Laboratory of Surface Physics, Fudan University, Shanghai 200433, China}
\affiliation{Collaborative Innovation Center of Advanced Microstructures, Nanjing 210093, China}

\author{C.~S.~Ting}
\affiliation{Texas Center for Superconductivity and Department of Physics, University of Houston, Houston, Texas 77204, USA}

\author{Yan~Chen}
\affiliation{Department of Physics and State Key Laboratory of Surface Physics, Fudan University, Shanghai 200433, China}
\affiliation{Collaborative Innovation Center of Advanced Microstructures, Nanjing 210093, China}

\date{\today}

\begin{abstract}

We study the \emph{orbital effects} of the synthetic magnetic fields in an interacting square lattice two-leg fermionic ladder model with the number-conserving pair hopping that hosting the Majorana bound states. By utilizing density matrix renormalization group and exact diagonalization, we identify a novel type-II topological Meissner $($\emph{topo}-Meissner$)$ phase $($as distinguished from the low-field type-I \emph{topo}-Meissner state$)$ when threading a \emph{high} magnetic flux through the plaquette of the ladder, which not only exhibits a large uniformly circulating chiral current along the legs, the characteristics of the celebrated Meissner state, but also accommodates a topologically protected ground-state manifold due to the \emph{reentrant} emergence of the Majorana end modes. Our work reveals some interesting interference effects resulting from the interplay between the gauge fields and the strong interactions in establishing the intrinsic topological states of matter in low-dimensional quantum systems.

\end{abstract}

\pacs{67.85.$-$d, 03.65.Vf, 74.25.Ha, 47.37.$+$q}

\maketitle

Topological superconductivity and superfluidity featuring the realization of Majorana bound states have comprised one of the major foci of condensed matter and cold atom communities in recent years \cite{Kitaev,Fu,Lutchyn,Oreg,Alicea,Elliott,Zhang,Sato,DalibardRMP}. One common thread in this pursuit entails interfacing materials with differing topological properties. Among the theoretical proposals, $1$D semiconducting nanowire models attract the great attention where due to the superconducting proximity effect and the strong spin-orbit coupling, unpaired Majorana zero modes could be materialized at the ends of the wire when applying the Zeeman fields \cite{Lutchyn,Oreg}. Interaction effects are incorporated later to verify the broad correctness of the mean-field predictions \cite{Stoudenmire}. In accordance with these theoretical breakthroughs, the experimental progress follows promptly that tentative signatures of Majoranas have been established \cite{Mourik,Das,Deng,Churchill,Higginbotham,NPerge}.

Generically, the solid-state architectures of Majorana fermions are in close contact with the external bulk superconducting and/or ferromagnetic materials, therefore the total number of particles is not a strictly conserved quantity in these systems. Moreover, the required strong proximity effects seem to render the experimental realization quite challenging \cite{Hart,Pribiag}. In comparison, ultracold atoms trapped in the optical lattices offer an alternative platform to engineer and simulate topological quantum matters in a relatively controllable and flexible manner \cite{Aidelsburger,AidelsburgerFlux,MiyakeFlux,AtalaZak,Atala,Jotzu,Mancini,Kennedy,AidelsburgerChernNum}. Particularly, most cold atom systems are particle-number conserved, and the parameters of the intrinsic interparticle interactions can also be tuned within a wide range \cite{BlochRMP}.

Currently the Zeeman effect of the applied magnetic fields is widely believed to play a vital role in the physics of Majorana fermions, which helps open a spin-orbit gap in the nanowire model that determines the size of the topologically nontrivial region \cite{Alicea}. An implementation of the \emph{phase} degrees of freedom of the gapping order parameters further brings about the possibility of realizing an exotic topological Fulde-Ferrell-Larkin-Ovchinnikov state in the Zeeman-field induced spin-imbalanced physical systems \cite{Chen,Qu,WZhang,Liu}. Despite these interesting developments, relatively little attention has been paid to the \emph{orbital effect} of the external magnetic fields, particularly its impacts on the realization of Majoranas in quasi-$1$D coupled-wire models. This situation gets somewhat reversed in the cold atom systems, where a great deal of effort has been devoted to creating strong synthetic magnetic fields in the optical lattices by using Raman-assisted tunneling, which mimics the standard Peierls substitutions through attaching an Aharanov--Bohm-like complex phase to the amplitudes of the nearest-neighbor single-particle hopping \cite{Aidelsburger,AidelsburgerFlux,MiyakeFlux,AtalaZak,Atala,Jotzu,Mancini,Kennedy,AidelsburgerChernNum}. The remarkable tunability of these artificial gauge fluxes has paved an experimental route to exploring the Bose-Einstein condensation in the Harper-Hofstadter model of the $2$D bosonic gases \cite{Kennedy}. Chiral edge states due to the cyclotron motion of the particles stirred by the flux have also been detected in the quasi-$1$D bosonic ladders \cite{Atala} and fermionic ribbons \cite{Mancini}.

Motivated by these recent experimental observations of the chiral currents of neutral fermions in the synthetic Hall ribbons \cite{Mancini}, in this Letter we study the phase diagrams of a minimal interacting two-leg ladder lattice model of spinless fermions that is coupled by the number-conserving pair hopping and exposed to the tunable artificial magnetic fluxes. Our most important finding based on the combined density matrix renormalization group $($DMRG$)$ \cite{White} and exact diagonalization $($ED$)$ \cite{Sandvik} simulation is the discovery of a new topological phase featuring a large uniform chiral current circulating along the boundaries of the ladder. In contrast to the celebrated Meissner state in the superconductors, this topological Meissner phase resides in the region of strong $($rather than weak$)$ gauge fluxes. To distinguish from the low-field phase of topological superfluid accompanied by a Meissner current $($type-I$)$, we call this new phase of matter the type-II \emph{topo}-Meissner state. Specifically, we find that when tuning the flux quanta per plaquette, the system first exits the type-I regime from the zero-flux point and subsequently enters a non-topological vortex phase under the moderate field strengthes. With the further increase of the magnetic flux, an intriguing reentrant phenomenon occurs: The system recovers its topological properties via the reemergence of a protected ground-state degeneracy, and simultaneously a noticeable surface current also starts to flow. In some sense, this novel Meissner current may be responsible for the reentrance into the newly-discovered type-II \emph{topo}-Meissner phase.

{\em Ladder model and current formulas.}---Here we consider the following number-conserving Hamiltonian for the spinless fermion lattice ladder model with $L$ rungs in the presence of a perpendicular $U(1)$ magnetic field,
\begin{align}
H\!=&-\!\sum^{L-2}_{j=0}\!\left[\left(t_{\parallel} e^{i\frac{\phi}{2}} c^{\dagger}_{j,0}c_{j+1,0}+t_{\parallel} e^{-i\frac{\phi}{2}} c^{\dagger}_{j,1}c_{j+1,1}\right)+\textrm{H.c.}\right] \nonumber \\
&+\!\sum^{L-2}_{j=0}\!\left(Wc^{\dagger}_{j,0}c^{\dagger}_{j+1,0}c_{j,1}c_{j+1,1}+\textrm{H.c.}\right) \nonumber \\
&-\!\sum^{L-1}_{j=0}\!\left(t_{\perp}c^{\dagger}_{j,0}c_{j,1}+\textrm{H.c.}\right),
\label{modelhamiltonian}
\end{align}
where $c^{(\dagger)}_{j,\ell=0,1}$ is the fermionic annihilation $($creation$)$ operator at rung $j$ on the leg $\ell=0,1$. The intraleg and interleg single-particle tunneling strengthes are $t_{\parallel}$ and $t_{\perp}$, respectively. Two essential ingredients of the above model are the synthetic Peierls phase $\phi\in[0,\pi]$ per plaquette that breaks the time-reversal symmetry, and the interchain pair-hopping interaction with the amplitude $W$ that preserves the fermion parity of one of the legs. In our numerical DMRG and ED simulations, we set $t_{\parallel}=1$ as the energy unit, and concentrate on the parametric regime of strong $W$ and small but nonzero $t_{\perp}$. Moreover, to faithfully characterize the topological ground-state manifold, we define the number parity operator for each leg: $(-1)^{N_{\ell}}$ where $N_{\ell}$ is total particle number operator of leg $\ell$. Previous works have demonstrated the realization of Majorana fermions in this Hamiltonian at $\phi=0$ \cite{Cheng,Sau,Kraus,Lang,Iemini}. In the present paper, we will show a richer phase diagram of Eq.~(\ref{modelhamiltonian}) when varying the flux from $0$ to $\pi$. Especially we make the prediction of a new type-II \emph{topo}-Meissner phase and study the associated topological quantum phase transitions.

Due to the insertion of the magnetic flux, there exist finite currents flowing along the legs and rungs of the ladder, which can be calculated via the Heisenberg equation: $J=\frac{\partial P}{\partial t}=i[H,P]$, where $P$ is the polarization operator. The obtained leg currents are
\begin{align}
\langle J^{\ell=0}_{j,j+1}\rangle&=2t_{\parallel}\!\cos\!\frac{\phi}{2}\Im\langle c^{\dagger}_{j,0}c_{j+1,0}\rangle+2t_{\parallel}\!\sin\!\frac{\phi}{2}\Re\langle c^{\dagger}_{j,0}c_{j+1,0}\rangle, \nonumber \\
\langle J^{\ell=1}_{j,j+1}\rangle&=2t_{\parallel}\!\cos\!\frac{\phi}{2}\Im\langle c^{\dagger}_{j,1}c_{j+1,1}\rangle-2t_{\parallel}\!\sin\!\frac{\phi}{2}\Re\langle c^{\dagger}_{j,1}c_{j+1,1}\rangle,
\end{align}
and the formula for the total rung current follows as per the Kirchhoff's conservation law of the particle currents,
\begin{align}
\langle J^{\perp}_{j}\rangle_{\textrm{tot}}&=\langle J^{\perp}_{j}\rangle_{\textrm{sing}}+\langle J^{\perp}_{j}\rangle_{\textrm{pair}}, \nonumber \\
\langle J^{\perp}_{j}\rangle_{\textrm{sing}}&=-2t_{\perp}\Im\langle c^{\dagger}_{j,0}c_{j,1}\rangle, \nonumber \\
\langle J^{\perp}_{j}\rangle_{\textrm{pair}}&=2W\Im\langle c^{\dagger}_{j,0}c^{\dagger}_{j+1,0}c_{j,1}c_{j+1,1}\rangle+(j\rightarrow j-1),
\end{align}
which contains contributions from both single-particle and pair tunnelings. Here we have chosen $\langle J^{\perp}_{j=-1,L}\rangle_{\textrm{pair}}=0$.

Before introducing the detailed picture of the type-II \emph{topo}-Meissner state and presenting the calculated phase diagrams, we would like to stress the significance of the sign of the pair-hopping amplitude $W$. A mean-field decomposition of the four-fermion term in Eq.~(\ref{modelhamiltonian}): $Wc^{\dagger}_{j,0}c^{\dagger}_{j+1,0}c_{j,1}c_{j+1,1}\rightarrow -W\Delta^{\dagger}_{p,\ell=0}\Delta_{p,\ell=1}$ where $\Delta_{p,\ell}=\langle c_{j,\ell}c_{j+1,\ell}\rangle$ suggests that when $W<0$, the $p$-wave superfluid order parameters in the two legs favor the out-of-phase configuration, however if $W>0$, the interchain superfluidity arrangement will tend to be in-phase. As will be shown in the next sections, this sign difference of $W$ has drastic consequences in determining the phase diagrams as tuning $\phi$: The type-II \emph{topo}-Meissner state could only be stabilized under the out-of-phase condition. Here we have assumed $t_{\perp}>0$.

{\em Type-II topo-Meissner state.}---The two salient properties that define the type-II \emph{topo}-Meissner phase are the following: (1) The existence of a topologically protected ground-state manifold that is spanned by two degenerate eigenstates with differing fermion parities due to the presence of Majorana end modes. (2) An accompanying uniform Meissner current circulates around the ladder in response to the \emph{strong} magnetic flux. A close inspection of Eq.~(\ref{modelhamiltonian}) suggests that these peculiar features stem from the nontrivial interplay among the single-particle tunneling, the pair-hopping interaction, and the $U(1)$ gauge field.

\begin{figure}[ht]
\centering
\includegraphics[width=0.48\textwidth]{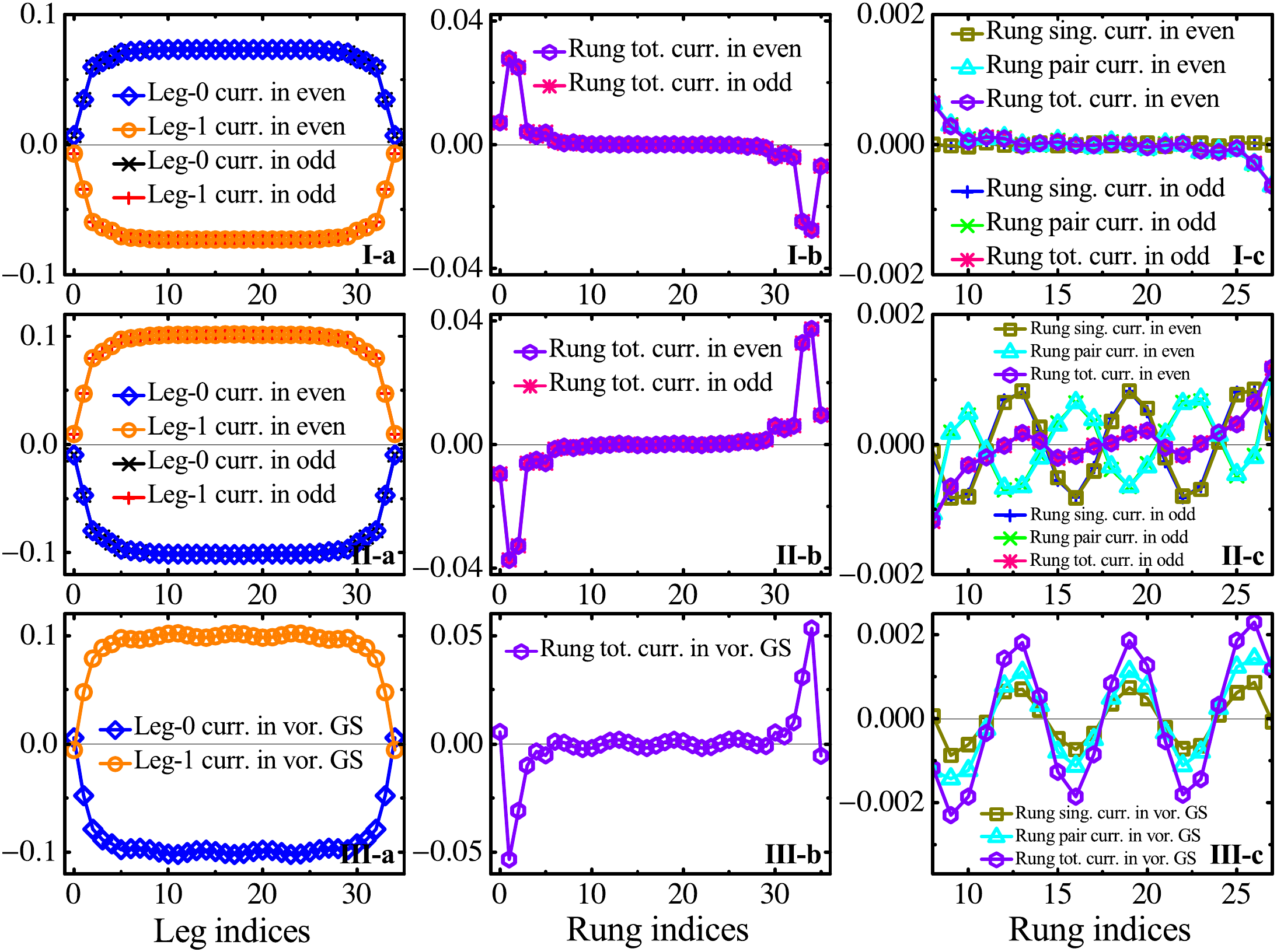}
\caption{\label{fig:fig1} (color online). Numerical DMRG results showing the leg and rung current patterns in the ladder model of $L=36$ rungs with the open boundary conditions. Panels II-a $($leg currents$)$, II-b $($rung currents$)$, and II-c $($rung currents in the bulk$)$ depict the type-II \emph{topo}-Meissner phase at $\phi=0.86\pi$ where a pronounced chiral current flows around the boundaries of the ladder demonstrating an unconventional Meissner-effect protected topological ground-state degeneracy due to the destructive interference between the rung single-particle current and the rung pair-hopping current, as illustrated in II-c. Here \lq\lq even'' and \lq\lq odd'' indicate the nonlocal fermion parity of one leg that distinguishes between the doubly degenerate eigenvectors in the manifold. Similarly, panels I-a, I-b, and I-c show the currents of the type-I \emph{topo}-Meissner state at low flux $\phi=0.1\pi$. Notice that the chiral current in the type-II Meissner state is reversed as compared to that in the type-I state. In the first two rows, we have fixed $t_{\perp}=0.1$, $W=-1.7$, and the total particle number $N=N_{\ell=0}+N_{\ell=1}=24$. Panels III-a, III-b, and III-c correspond to the non-degenerate ground state of the system in the non-topological vortex phase realized by switching the sign of the pair hopping by setting $W=1.7$ at the same high-flux point $\phi=0.86\pi$ where due to the constructive interference, the vortex state is facilitated while the topological state is suppressed, as can be seen from III-c.}
\end{figure}

The first two rows of Fig.~\ref{fig:fig1} display the prototypical leg and rung current configurations of the topologically nontrivial Meissner states obtained from the DMRG calculation. At \emph{low} magnetic fields $(\phi=0.1\pi)$, in view of the insensitivity of Majorana end modes to the \emph{weak} time-reversal-symmetry breaking perturbations, we expect a topological state with strong superfluid correlations that carrying a responsive leg current. This is shown in the panels I-a, I-b, and I-c where both rung currents $(\langle J^{\perp}_j\rangle_{\textrm{sing}}$ and $\langle J^{\perp}_j\rangle_{\textrm{pair}})$ of the type-I \emph{topo}-Meissner state are vanishingly small in the bulk region $($see I-c$)$. However, when the gauge flux $\phi$ increases to a pretty large value of $0.86\pi$, a novel \emph{destructive} interference effect appears: Even though both $\langle J^{\perp}_j\rangle_{\textrm{sing}}$ and $\langle J^{\perp}_j\rangle_{\textrm{pair}}$ develop the vortex structures, their relative oscillations are out-of-phase $($see II-c$)$, which gives rise to the cancelation of the rung currents in the bulk and the circulation of a uniform chiral current along the edges of the ladder. The establishment of this Meissner current is crucial for the realization of the type-II \emph{topo}-Meissner phase and the system's recovery of Majorana bound states at such high fluxes. The revealed destructive mechanism due to the peculiar response of the pair-tunneling interaction to the strong magnetic field also underpins the differentiation between the two types of \emph{topo}-Meissner states. Furthermore, it is well-known that the topologically degenerate ground states cannot be discriminated by any local physical quantity measurements. This point is manifested in our numerical results: The parity-even state shares the same lowest eigenenergy and the same current as well as particle-density profiles with the parity-odd state. The featured nonlocal distinction in the fermion parity makes the type-II \emph{topo}-Meissner phase robust $($at least in principle$)$ against local perturbations.

The second signature that distinguishes the type-II \emph{topo}-Meissner state from the type-I state is the chiral-current \emph{reversal}, as can be seen from Figs.~\ref{fig:fig1} I-a, II-a and I-b, II-b. Similar phenomena have been reported recently in bosonic ladders with interactions, we thus speculate that this observed switch in current chirality may also be due to the spontaneous enlargement of the lattice unit cell that leads to an increase of the effective magnetic flux \cite{Greschner}.

In stark contrast, as we change the value of $W$ from $-1.7$ to $1.7$ while keeping all other parameters intact in the high-flux region $(\phi=0.86\pi)$, the ladder system collapses into a non-topological vortex state: Both the total leg and rung currents are spatially modulated and their patterns show the vortex-like oscillations in the bulk region, as can be seen from Figs.~\ref{fig:fig1} III-a and III-b. More importantly, the excitation gap between the two lowest-lying energy levels becomes finite. The disappearance of the ground-state degeneracy in this case could be understood by noticing the resultant \emph{constructive} interference between $\langle J^{\perp}_j\rangle_{\textrm{sing}}$ and $\langle J^{\perp}_j\rangle_{\textrm{pair}}$ at $W=1.7$ $($see III-c$)$. This in-phase configuration enhances the oscillations of $\langle J^{\perp}_j\rangle_{\textrm{tot}}$, which promotes the vortex formation. Since vortices create gapless excitations in the bulk that shrink the spin gap, the Majorana end modes would no longer be protected.

\begin{figure}[ht]
\centering
\includegraphics[width=0.43\textwidth]{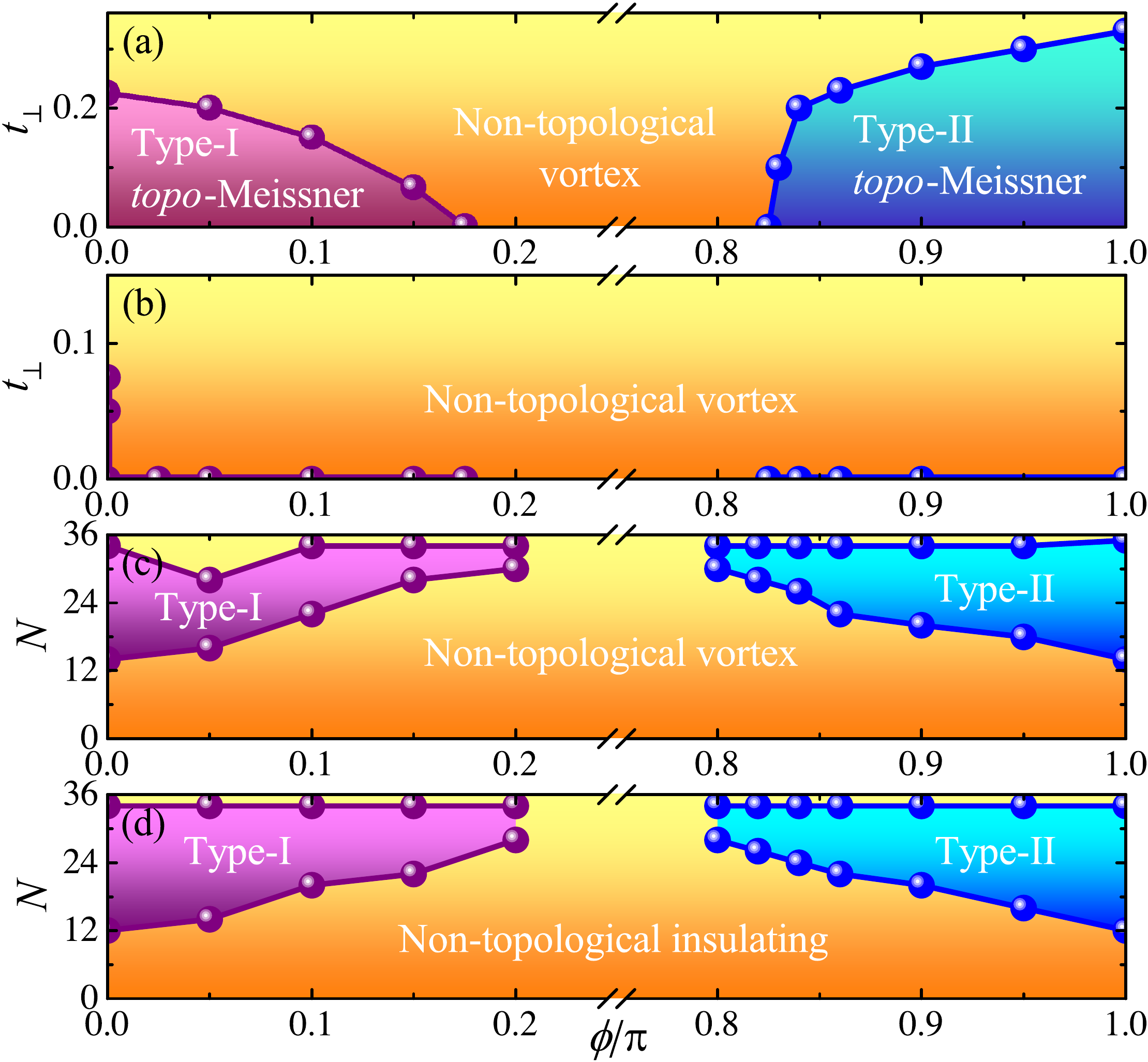}
\caption{\label{fig:fig2} (color online). Phase diagrams of the ladder model (\ref{modelhamiltonian}) in the plane of $t_{\perp}$ versus $\phi$ [(a),(b)] and $N$ versus $\phi$ [(c),(d)], where $N=24,~W=-1.7$ in (a), $N=24,~W=1.7$ in (b), $t_{\perp}=0.1,~W=-1.7$ in (c), and $t_{\perp}=0.0,~W=\pm1.7$ in (d). The ladder size $L$ is $36$.}
\end{figure}

{\em Phase diagrams.}---To achieve a better understanding of the Hamiltonian, we perform an extended DMRG calculation to map out the related phase diagrams. Figs.~\ref{fig:fig2}(a) and (b) plot the varied possible phases of (\ref{modelhamiltonian}) in the coordinates of $t_{\perp}$ versus $\phi$ at $W=-1.7$ and $1.7$, respectively. As advertised, the sign change in $W$ gives rise to very different results: Both types of \emph{topo}-Meissner states can only be stabilized at finite $t_{\perp}$ when $W$ is negative. In particular, there exist three distinct regions in panel (a): Type-I and type-II \emph{topo}-Meissner phases reside in the vicinity of point $\phi=0$ and $\phi=\pi$, respectively, which are separated by a non-topological vortex state in the middle range. Therefore through tuning the synthetic flux, we find a reentrant phenomenon of topological superfluidity. On the contrary, when $W=1.7$, the topologically nontrivial phases become fragile that give place to the non-topological vortex state if including tiny parity-breaking $t_{\perp}$ and/or finite flux $\phi$, as shown by panel (b). For completeness, the phase diagrams of $N$ versus $\phi$ are presented in panels (c) and (d) \cite{Footnote}.

\begin{figure}[ht]
\centering
\includegraphics[width=0.43\textwidth]{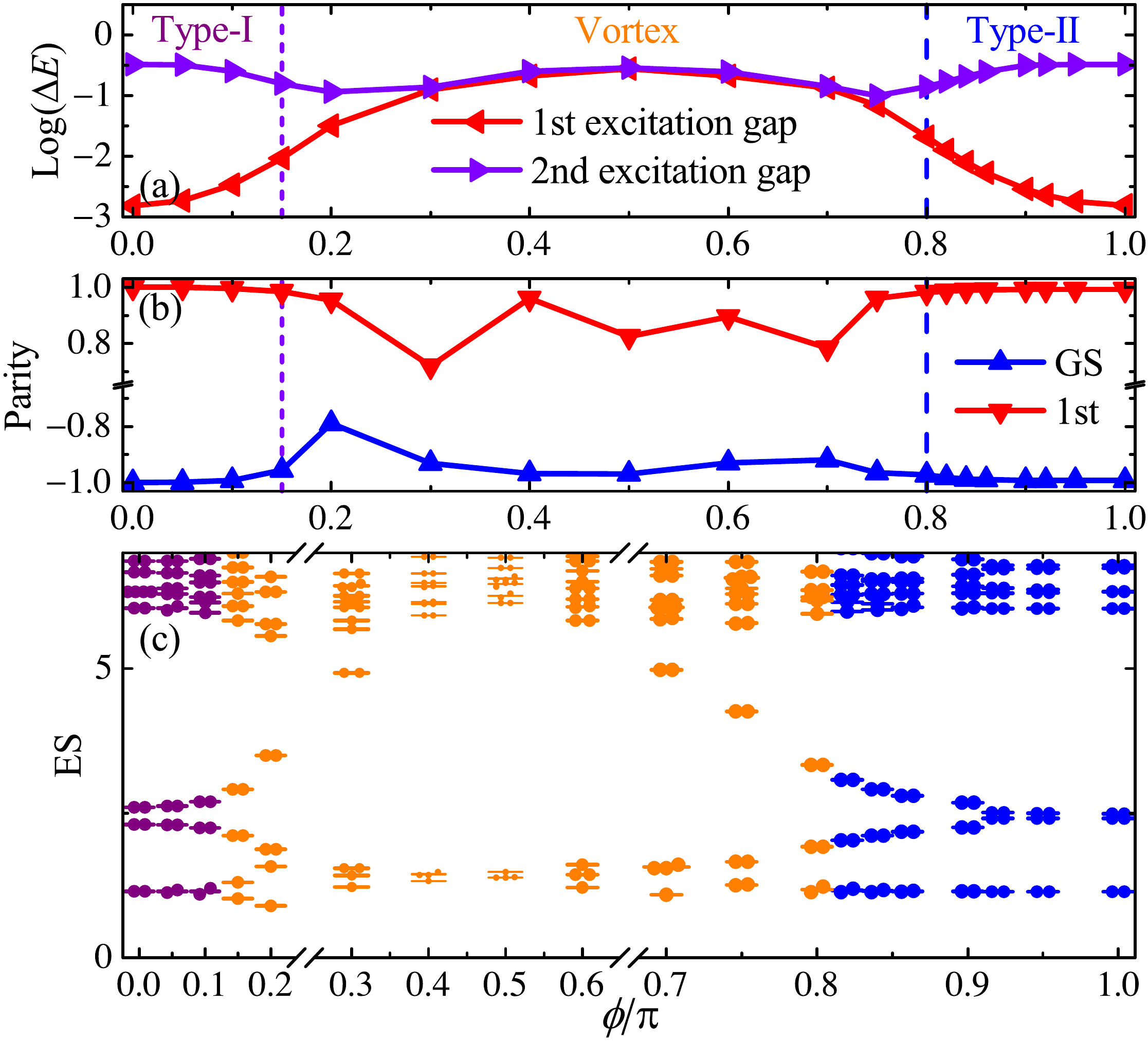}
\caption{\label{fig:fig3} (color online). ED results for the evolutions of (a) the excitation gaps, (b) the fermion parities, and (c) the entanglement spectra $($ES$)$ as functions of the tunable flux $\phi$. Here we have set the system's size $L=14$, the filling $N=10$, $W=-1.7$, and $t_{\perp}=0.1$.}
\end{figure}

{\em Topological quantum phase transitions.}---To unbiasedly pin down the reentrant topological transitions from type-I to type-II Meissner states, we solve the Hamiltonian (\ref{modelhamiltonian}) on a small ladder $(L=14)$ with $N=10$ fermions by an ED calculation. The obtained results are given by Fig.~\ref{fig:fig3} which confirm and complement the observations in the DMRG simulation. In particular, Figs.~\ref{fig:fig3}(a) and (b) plot the evolutions of the first two excitation gaps and the fermion parities of the two lowest-lying eigenstates as we tuning the flux $\phi$ from $0$ to $\pi$. As shown in the middle $($non-topological vortex$)$ region, there is a unique ground state that is separated from the nearly degenerate $1$st and $2$nd excitation states by a noticeable energy gap. The corresponding fermion parities of these energy levels also show a deviation from $+1$ $($even$)$ and $-1$ $($odd$)$ in the center region indicating an enhanced parity-breaking effect of $t_{\perp}$ $(=0.1)$ driven by the magnetic fields. However, when the flux $\phi$ exceeds a critical value $(\sim\!0.8\pi)$, the $1$st excitation gap of the system begins to quickly decrease while the $2$nd excitation gap is slightly increased, which leads to the reemergence of a degenerate ground-state manifold that can be spanned by two well-defined parity-even and parity-odd sectors [see panel (b)]. This demonstrates the operation of the proposed destructive interference mechanism in the high-field type-II region.

Furthermore, the many-body Majorana end modes can be visualized via the bulk entanglement spectrum, the lowest part of which is supposed to be doubly degenerate if the system is in the topologically nontrivial phases \cite{Stoudenmire,Li,Pollmann}. The calculated reduced density-matrix eigenvalues at the central bond are depicted in Fig.~\ref{fig:fig3}(c), from which we can perceive the following: (1) The twofold entanglement degeneracy is only present in the type-I and type-II regions. (2) The evolution of the gaps in the entanglement spectrum shows a closing-reopening-closing-reopening pattern, which mimics the behavior of the spin $($antibonding$)$ gap in the two successive topological phase transitions when gradually increasing $\phi$. (3) The appearance of the twofold degeneracy in the lowest entanglement spectrum coincides with the recovery of well-defined ground-state parities in the transition from vortex state to the type-II \emph{topo}-Meissner state. These points are consistent with the excitation-gap and parity measurements in panels (a) and (b). Combining all the above analyses we could conclude the existence of two types of \emph{topo}-Meissner states in model (\ref{modelhamiltonian}), and it appears that the type-II phase is more stable than the type-I phase in the presence of finite $t_{\perp}$ and $\phi$.

{\em Experimental considerations.}---Fermionic chiral edge states in a Hall ribbon have been experimentally detected recently, where both the magnetic fields and one of the spatial dimensions are synthetic \cite{Mancini}. The required strong artificial gauge fluxes can be generated by either the Raman laser beams or the techniques of lattice modulation \cite{Struck}. In Ref.~\cite{Kraus} an atomic scheme has also been outlined for realizing the more complicated pair-hopping interactions with suppressed single-particle tunnelings. Therefore a reasonable part of our results might be testable in the lab.

In summary, our numerical simulation suggests the existence of a topologically nontrivial type-II Meissner state in the high-field region of the pair-hopping-coupled atomic Fermi wires. This exotic phase of matter becomes possible due to the destructive interference between the interleg single-particle and pair-hopping currents driven by the strong artificial gauge flux. The associated generic phase diagrams of model (\ref{modelhamiltonian}) have been obtained in a large-size lattice by using DMRG and the involved topological quantum phase transitions have been confirmed in a small ladder by using the unbiased ED. Since the requisite ingredients of our proposal could be designed and engineered in the cold-atom laboratory, we expect the potential realization and detection of the predicted type-II \emph{topo}-Meissner phase in the foreseeable future.

{\em Acknowledgments}: C.~C. is grateful to Prof. F.~J.~Burnell for the careful reading of the manuscript and for constructive suggestions and questions. C.~C. also thanks M.~D.~Schulz for useful discussions. Part of the calculation is developed from ALPS package \cite{Dolfi}. W.~Y. and Y.~C. are supported by the State Key Programs of China (Grant No.~$2012$CB$921604$), the National Natural Science Foundation of China (Grant Nos.~$11274069$ and $11474064$). C.~S.~T. has been supported by the Robert A. Welch Foundation under Grant No.~E-$1146$.

C.~C. and W.~Y. contributed equally to this work.


\begin{thebibliography}{99}
\bibitem{Footnote} We find that at the half and quarter fillings, the system is typically non-topological due to the backscattering.
\end{thebibliography}
\end{document}